# TurPy: a physics-based and differentiable optical turbulence simulator for algorithmic development and system optimization

Joseph L. Greene[a,*], Alfred Moore[a], Iris Ochoa[a], Emily Kwan[a], Patrick Marano[a], and Christopher R. Valenta[a]
[a]Georgia Tech Research Institute, Atlanta, GA, 30318, USA
*Joseph.Greene@gtri.gatech.edu

## ABSTRACT

Developing optical systems for free-space applications requires simulation tools that accurately capture turbulence-induced wavefront distortions and support gradient-based optimization. Here we introduce TurPy, a GPU-accelerated, fully differentiable wave optics turbulence simulator to bridge high fidelity simulation with end-to-end optical system design. TurPy incorporates subharmonic phase screen generation, autoregressive temporal evolution, and an automated screen placement routine balancing Fourier aliasing constraints and weak-turbulence approximations into a unified, user-ready framework. Because TurPy's phase screen generation is parameterized through a media-specific power spectral density, the framework extends to atmospheric, oceanic, and biological propagation environments with minimal modification. We validate TurPy against established atmospheric turbulence theory by matching 2nd order Gaussian beam broadening and 4th order plane wave scintillation to closed-form models with 98% accuracy across weak to strong turbulence regimes, requiring only the medium's refractive index structure constant and power spectral density as inputs. To demonstrate TurPy as a gradient-based training platform, we optimize a dual-domain diffractive deep neural network (D2NN) in a two-mask dual-domain architecture to recover a Gaussian beam from a weakly turbulent path and achieving over 58% reduction in scintillation relative to an uncompensated receiver in simulation. TurPy is released as an open-source package to support synthetic data generation, turbulence-informed algorithm development, and the end-to-end design of optical platforms operating in turbulent environments.

## 1. INTRODUCTION

As data-driven approaches to optical design and algorithm development enable increasingly capable free-space optical systems there is a growing demand for simulation tools that accurately reproduce physical phenomena to support training in data-limited regimes. This challenge is pronounced for systems operating through optical turbulence, which induces medium-dependent refractive index fluctuations that distort, break up, and refract propagating beams as a function of wavelength, beam profile, and instantaneous environmental conditions [1]. As collecting field data across the full range of beam configurations, environmental states, and propagation geometries is impractical, the development of high-fidelity simulators serves as a priority for robust optical technologies.

Time-averaged simulators characterize aggregate effects, including beam broadening and scintillation, through closed-form solutions for select media [2], [3] and applications [4], [5], [6], but fail to capture the instantaneous wavefront distortions that govern signal degradation at the cost of simplicity. Numerical wave optics simulators recover that information by modeling field distributions through stochastic media directly, but at substantially greater computational cost. Split-step methods balance this tradeoff by decoupling propagation from turbulence-induced error, alternating between a diffraction step governed by a standard optical transfer function and a real-space modulation by a statistically representative phase screen [7]. In their basic form, split-step methods must assume weak turbulence, spatial stationarity, finite aperture, isoplanatic conditions, and coherent illumination to remain physically valid. Subsequent work has relaxed individual assumptions to incorporate subharmonics [8], strong turbulence [9], temporal correlation [10], anisoplanatism [11], and partial coherence [12], improving fidelity while retaining computational efficiency. As a result, split-step simulators have served as effective synthetic testbeds for algorithm development [13] and wavefront engineering, including structured light [7] and adaptive optics [14]. However, emerging deep optics frameworks, which jointly optimize optical elements and downstream algorithms end-to-end in gradient-based frameworks, place additional demands on simulator speed, differentiability, and physical fidelity that conventional split-step tools were not designed to meet [15], [16], [17]. As a result, no existing tool, to our knowledge, consolidates anisoplanatism, subharmonic generation, temporal correlation, and strong turbulence models within a GPU-accelerated, differentiable architecture suitable for end-to-end optimization (see **Appendix A**).

This work introduces TurPy, a wave optics turbulence simulator built for differentiable, end-to-end system optimization. TurPy generates physically accurate phase screens from user-defined turbulence profiles, while enforcing both Fourier aliasing constraints and weak-turbulence approximations through an automated screen placement routine, and supports GPU-accelerated autoregressive temporal evolution, anisoplanatism, and subharmonic generation. Additionally, TurPy preserves efficient gradient flow by basing its split-step simulation in PyTorch to enable automated gradient tracking and application to optimize upstream optical design. Because TurPy's phase screen generation is parameterized through a media-specific power spectral density (PSD), the framework extends to atmospheric, oceanic, and biological propagation environments with minimal modification (see **Appendix B**).

To validate TurPy as a physically accurate tool, we simulate optical fields over independent realizations of atmospheric turbulence and compare their aggregate statistics to known theory. Specifically, we match 2nd order (e.g., average) broadening of a Gaussian beam and 4th order (e.g., variance) scintillation of a plane wave with 98% accuracy when compared to theory. We highlight that achieving this close agreement solely requires information on the index of refraction and PSD of the target media to empirically describe these stochastic effects. To demonstrate TurPy as a gradient-based training platform, we use it to optimize a diffractive deep neural network (D2NN), which are an engineered sequence of static phase surfaces (here called phase masks – not be confused with split step phase screens), requiring no powered components or feedback loops to achieve leading wavefront control all-optically and passively when applied to image processing [18], optical encryption [19], and statistical inference [20]. Here, we train a two-mask dual-domain architecture to recover a Gaussian beam from a weakly turbulent path and achieve a 58% reduction in scintillation relative to an uncompensated receiver in simulation. **Section 2** presents TurPy's theoretical framework. **Section 3** validates TurPy against established atmospheric statistics across weak to strong turbulence regimes, and **Section 4** demonstrates the D2NN training application.

## 2. METHODS

### 2.1 Representing Optical Turbulence through a Split-Step Model

Simulating optical propagation through turbulence requires decoupling the deterministic diffraction of the field from the stochastic phase accumulation induced by refractive index fluctuations. Split-step methods accomplish this by using weakly scattering approximations (e.g., Rytov, First-Born) to decouple optical propagation from error accumulation to use standard optical models (e.g., Huygen-Fresnel) to describe propagation through turbulent media. In this formulation, the split-step approach operates on an optical field in two steps.

First, an input complex optical field $U(x,y) = A(x,y)e^{i\phi(x,y)}$ is defined where $(x,y)$ are cartesian coordinates, $A(x,y)$ is the spatial amplitude factor and $\phi(x,y)$ is the spatial phase factor, and modulated by a Fourier space propagation transfer function, $H(\kappa_x, \kappa_y; \Delta z)$ where $\kappa_x, \kappa_y$ are cartesian wavevectors and $\Delta z$ is the propagation step size. Next, the propagated field is modulation by a real-space statistical phase screen, $\varphi_n(x,y;\ \Delta z)$, to yield a step-step propagation step of:

$$U_{\Delta z}(x,y) = \mathfrak{I}^{-1}\left\{\mathfrak{I}\{U(x,y)\} \times H\left(\kappa_x, \kappa_y, \Delta z\right)\right\} \times e^{i\varphi_n(x,y;\Delta z)} \qquad 1$$

where $U_{\Delta z}(x,y)$ is the propagated field and $\mathfrak{I}$, $\mathfrak{I}^{-1}$ are Fourier and inverse discrete Fourier transforms (DFTs), respectively. In this work, we will use an angular spectrum propagation transfer function versus a Fresnel approximation to accurately model the high degree of angular multiplexing induced by turbulence along highly turbulent paths.

To generate a representative phase screen, the power spectral density (PSD) of optical turbulence (e.g., a description of how fluctuating energy is distributed across spatial scales) is used and observed to follow media-specific power laws within inertial regimes. Using this power law description, the spatial distribution of phase screens can be characterized and subsequently scaled by the projected phase accumulation from index variation across an interval. The full phase screen derivation may be found in **Appendix B**.

TurPy uses this description to create a phase screen generation pipeline that builds on user-defined parameters defining media and turbulence over a path (see **Fig. 1**). To initialize a turbulence profile, TurPy takes in the desired range intervals between split steps, a description of optical turbulence strength along the path (here parameterized by the index of refraction structure constant, $C_n^2$ – a term to describe optical turbulence strength within the inertial subrange over distance) and user-defined constants specifying the simulation, such as average index, inertial subrange scale limits, and simulated grid size (see **Fig. 1A**). Here, the subrange bounds when the PSD power law breaks down at an inner scale, $l_0$ (i.e., due to viscous forces), and outer scale, $L_0$ (i.e., due to convective forces). Next, TurPy calculates turbulence strength across an interval. For the atmospheric examples, TurPy uses the Fried Parameter, $r_0$, as a path-integrated quantity of spatial coherence. After calculation, the PSD is generated on a synthetic grid and modulated with a complex normal Gaussian random noise (see **Fig. 1B**). These Gaussian examples are used to create statistically consistent yet instantaneously randomized phase variations with amplitudes that follow the desired distribution. Next, the phase screens at each interval are converted from their k-space PSD representation (here denoted by $\kappa_{\perp}$) to real space (here denoted by $r_{\perp}$) through a discrete Fourier transform (DFT) to create a physical realization for use in **Eq. 1** (see **Fig. 1C**). Finally, an optical field is propagated through each interval to create a synthetically perturbed beam. An example of a Gaussian beam (see **Fig. 1D**) propagated over a 500 m path is shown with turbulence (here, $\sigma_R^2 = 0.1$) and without turbulence.

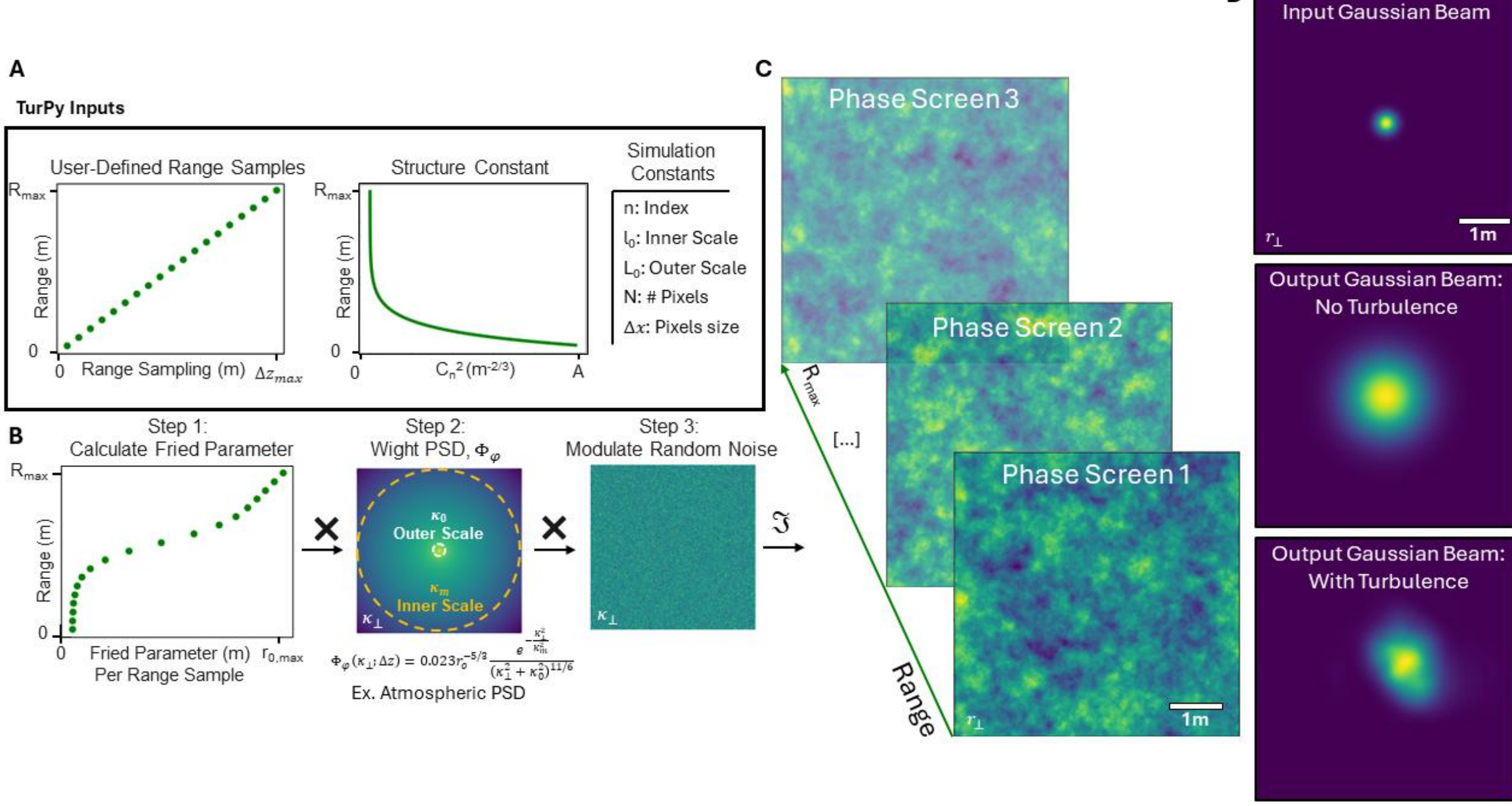

**Fig. 1. Overview of Phase Screen Generation.** **(A)** User inputs desired range samples, turbulence strength over path, as well as media and simulation parameters. **(B)** TurPy calculates Fried parameter and weights PSD with random modulation to create statistically consistent realizations. **(C)** Phase screens are generated at each interval along the path with variance defined by interval-integrated turbulence strength. **(D)** TurPy may be used with (bottom) and without (middle) turbulence as a free space propagator for an input signal (top).

## 2.2 Subharmonics for Outer Scale Effects

Often, simulated grids exhibit a low number of pixels to improve the computational efficiency of wave optics simulations, due to the O(NlogN) efficiency of the DFT. However, this design choice implicitly sets the largest scale of phase variations captured within a phase screen, which may lead to physical inaccuracies if not set appropriately. This effect becomes prominent in applications such as atmospheric turbulence, where the outer scale of turbulence may range from 1-100m dependent on altitude and wind speed [4], while wave optics propagation demands optically relevant discretization. To generate outer scale effects without greatly inflating array size, TurPy leverages subharmonic generation methods [8], which serves as a recursive method to generate a phase screen correction factor on sub-grids near the DC frequency coordinate to capture low-frequency information below the nominal Fourier space resolution.

To generate a subharmonic phase screen, we first define a subharmonic recursion level, $N_p$, and subharmonic scale factor, $s$. The recursion level determines the number of sub-grid levels we generate and the subharmonic scale determines the effective resolution increase per sub-grid. The resulting subharmonic phase screen will effectively exhibit $s^{N_p}$ finer resolution. $s$ is assumed to be odd (here, 3) to preserve a (0,0) DC coordinate and this work assumes 3 sub-grid levels for a 27x increase in resolution across all simulations. Once the subharmonic parameters are set, the subharmonic phase screen is generated through:

$$\varphi_{SH}(x,y;\Delta z) = \sum_{p}^{N_p} \sum_{m,n=-\frac{s-1}{2}}^{\frac{s-1}{2}} \sqrt{\Phi_\varphi\left(\frac{\kappa_x}{s^p},\frac{\kappa_y}{s^p};\Delta z\right)\Delta\kappa_x\Delta\kappa_y} \times \left(\mathrm{N_c}\left(0,1;N_x,N_y\right)\right)\} \times e^{\frac{i2\pi}{s^p}\left(\frac{n}{N_x}x+\frac{m}{N_y}y\right)} \tag{2}$$

Where $\varphi_{SH}(x,y;\Delta z)$ is a subharmonic phase screen generated on a cartesian (x, y) grid over interval $\Delta z$, $\Phi_\varphi\left(\frac{\kappa_x}{s^p},\frac{\kappa_y}{s^p};\Delta z\right)$ is the path-integrated phase surface PSD over spatial frequencies $(\kappa_x,\kappa_x)$, $\Delta\kappa_x,\Delta\kappa_y$ are the k-space resolutions used to preserve power in accordance with Parseval's theorem, and $\mathrm{N_c}\left(0,1;N_x,N_y\right)$is a normal Gaussian random variable with $\mu=0,\sigma=1$ generated on a grid of size $N_x\times N_y$.

Functionally, **Eq. 2** is a truncated inverse discrete Fourier transform defined manually over the captureed sub-grid frequencies. As shown in **Fig. 2**, this procedure captures increasingly finer PSD features into the subharmonic screen to include low-order (e.g., tip-tilt) features as if arising from outside the original array size (see **Fig. 2C**). Each intermediate subharmonic phase screen is added to the original phase screen in real space through:

$$\varphi_{final}(x,y;\Delta z) = \varphi_n(x,y;\Delta z) + \varphi_{SH}(x,y;\Delta z) \tag{3}$$

Due to only requiring a small number of coefficients and a truncated inverse DFT to compute, the subharmonic method provides an efficient alternative for capturing outer scale effects into simulated turbulence without sacrificing efficiency.

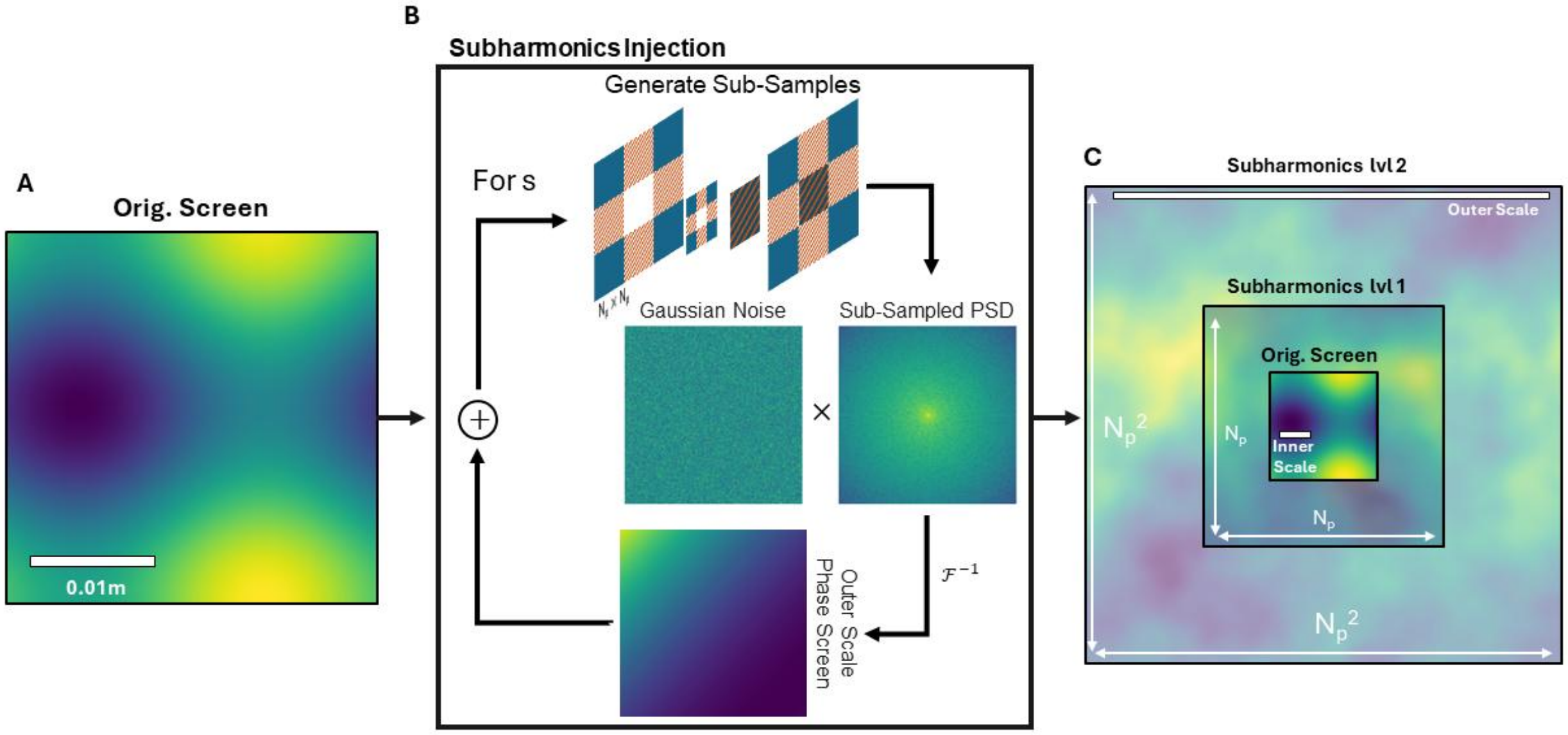

**Fig. 2 Subharmonic Phase Screen Generation. (A)** An original screen on the order of an inner scale is passed in. **(B)** The subharmonic generation routine generates subsampled PSD information to add global effects. **(C)** the resulting screen is appended such that it appears as a representative patch of a larger screen.

### 2.3 Temporal Correlation through Autoregression

Up until this point, phase screens between split-step iterations are assumed to be uncorrelated. However, in fast paced application with dynamic environments, such as free space optical communication (FSOC), optical signatures may repeat at a frequency higher than the Greenwood frequency (e.g. the rate at which turbulent patches change). This effect leads to the need to simulate temporally correlated features. In general, turbulence evolves with two effects: by translation, through effects like winds or currents, and by reorganization of turbulent structures, through effects like vapor pressure or convective forces. While simulators may account solely for the former for simplicity (see **Appendix A**), here TurPy leverages an autoregression-based approach to account for both mechanisms based on [10]. To account for translation effects, TurPy introduces a Fourier space transfer function using phase tilt, $\theta$, relating to a 2D translation velocity, $(v_X, v_y)$, and the inter-simulation evolution period, $T$:

$$e^{i\theta(f_x,f_y)} = e^{-2\pi T\left(f_x v_x + f_y v_y\right)} \quad (4)$$

To efficiently track our phase screen generation, while supporting access to the Fourier domain, TurPy stores the phase screen Fourier spectrum $\widetilde{\varphi_n}(\kappa_x, \kappa_y)$ between iterations. As such, TurPy may translate the phase screen by multiplying the Fourier representation by **Eq. 12** before taking the DFT.

This version of translation leads to circulant convolution artifacts, requiring captureion of new information across the phase screen to not periodically repeats features. To enforce decorrelation, TurPy uses a phase attenuation factor, $\alpha$, which generates a new scaled PSD-weighted white Gaussian variable and adds it to the original phase screen spectrum to enforce decorrelation through:

$$\varphi_n(x,y;\Delta z) = \Im^{-1}\{\alpha\widetilde{\varphi_n}\left(f_x,f_y\right) + \sqrt{1-\alpha^2}\sqrt{\Phi_\varphi\left(\kappa_x,\kappa_y;\Delta z\right)\Delta\kappa_x\Delta\kappa_y} \times \mathrm{N_c}\left(0,1;N_x,N_y\right)) \quad (5)$$

Where $\sqrt{1-\alpha^2}$ is used to stabilize the total variance of the phase screen during evolution. As demonstrated below (see **Fig. 3**), this technique implicitly relates $\alpha$ to a decorrelation time. By relating $\alpha$ to a decorrelation time, TurPy may capture realistic decorrelation into our simulations using efficient operations (e.g., multiplying the constant time transform function and adding the decorrelation PSD) in a preserved phase screen spectrum.

The autoregression is shown in **Fig. 3**. First, an original phase screen is defined with an assumed wind vector. The phase tilt is calculated and a new random phase screen is generated. The original phase screen is displaced by the phase tilt and weighted by the new random sample using the blending factor, $\alpha$, to generate a new screen, as shown in **Fig. 3A**. Over short intervals and high $\alpha$, the phase screen features remain correlated as displacement dominates, however, over long time the phase screen realization drifts due to the captureion of new random noise, as shown in **Fig. 3B**. To characterize how $\alpha$ influences correlation time, phase screens are simulated over 30 fixed random seeds with no displacement and their Pearson's cross correlation score is monitored over 250 iterations, as shown in **Fig. 3C**. A Pearson's cross correlation score of 0.4 is assumed as sufficient decorrelation and used to fit fit the decorrelation time to $\alpha$ as a one-time calibration, as shown in **Fig. 3D**. After calibration, this trend can be used to estimation decorrelation time of our media and set TurPy appropriately.

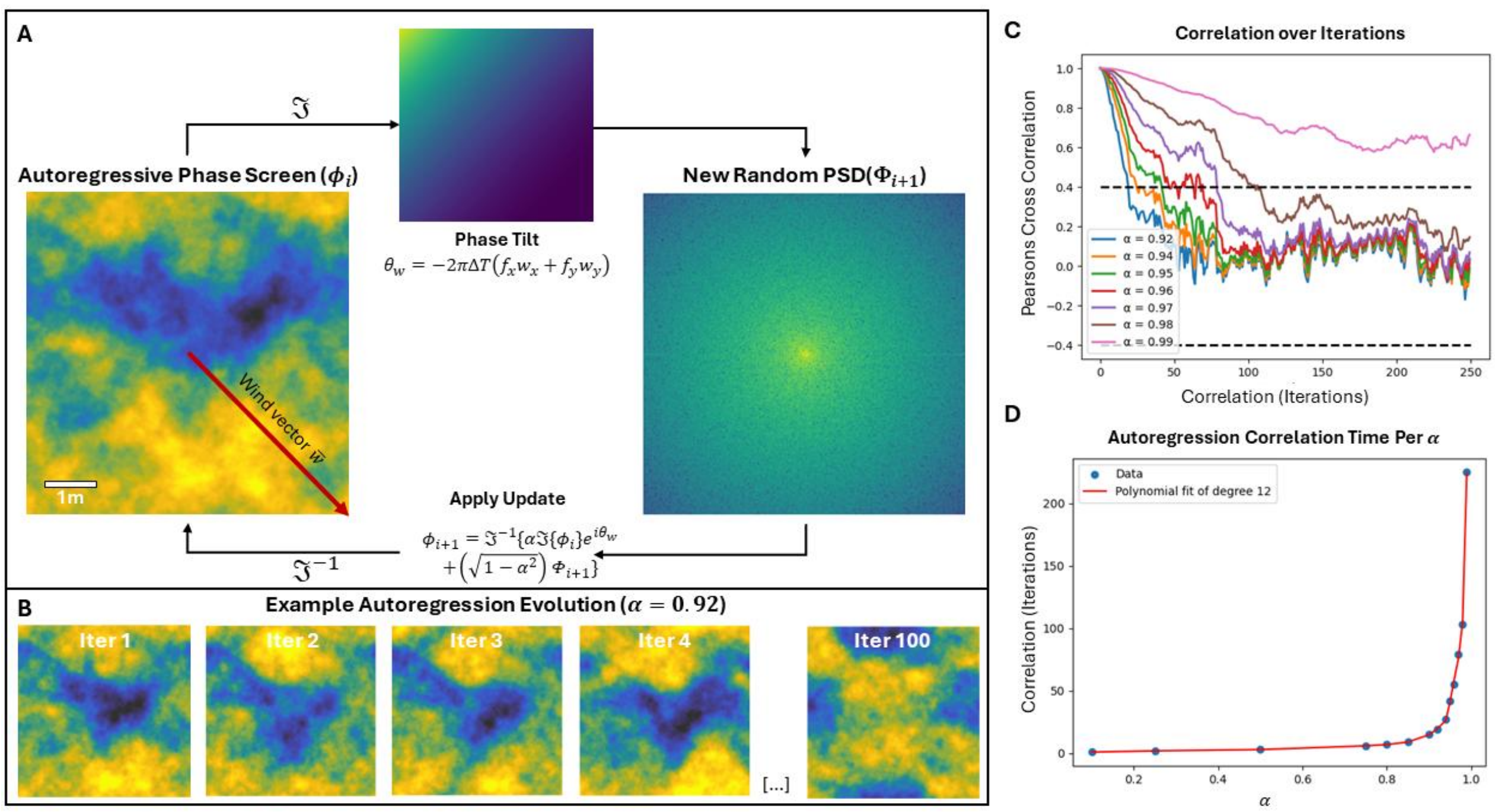

**Fig. 3. Autoregression for Phase Screen Evolution.** **(A)** Autoregression algorithm of Phase Screen using Fourier space tilt for translation and random PSD captureion for evolution. **(B)** Example of short time and long-time evolution of a phase screen. **(C)** Calibrated decorrelation over $\alpha$ with **(D)** polynomial fit.

## 2.4 Phase Screen Placement Optimization to Balance Fourier Constraints & Phase Screen Variance

Split-step simulations must balance two constraints to maintain physical accuracy: Fourier space aliasing and weakly scattering approximations. To maintain the first constraint, Fourier optics demands that an angular spectrum transfer function may not induce a phase gradient that aliases when simulated on a discretized grid. To uphold this requirement, the phase gradient of the transfer function is determined as a function of stepsize.

Here, the angular spectrum transfer function in k-space takes the form:

$$H_{as}(\kappa_x, \kappa_y; \Delta z) = \exp\left(i\Delta z\sqrt{n^2 k_0^2 - \kappa_x^2 - \kappa_y^2}\right)$$

Where $H_{as}$ is the angular spectrum transfer function, $\Delta z$ is the stepsize, $n$ is the average refractive index, and $k_0 = \frac{2\pi}{\lambda}$ is the wavenumber in vacuum. Next, we take the gradient of the phase argument (e.g., $\Psi_0 = \frac{\partial \arg(H_{as})}{\partial \kappa_0}$) and solve it at the maximum k-space coordinate supported by the simulation (e.g., $\kappa_x^2 + \kappa_y^2 = \kappa_0^2 = \left(\frac{2\pi}{\Delta x}\right)^2$) and enforcing the Nyquist criteria $\Delta\kappa \leq \frac{1}{2\Psi_x}$. Here, $\Delta\kappa$ represents the k-space resolution and it determined by the DFT reciprocal relationship, $\Delta\kappa = \frac{1}{N\Delta x}$.

Solving for $\Delta z$ yields:

$$\Delta z = \frac{N\Delta x}{2} * \left| \frac{\kappa_0}{\sqrt{n^2 k_0^2 - \kappa_0^2}} \right| \tag{7}$$

Where here $\Delta z$ is the maximum axial step size supported by the angular spectrum transfer function before aliasing. Users may increase the axial step size by a number of pixels (which trades computational efficiency) or by increasing $\Delta x$ (which trades spatial resolution of the simulated optical fields).

However, phase screen simulations additionally must support a weak approximation to decouple phase error from propagation. Generally, this condition is supported when Rytov variance ($\sigma_R^2$, e.g., a path integrated quantity of turbulence strength – see **Appendix C**) is less than 1. When this condition is violated, strong effects, such as decorrelation and chaotic scattering, may be under-expressed.

To address this need, TurPy incorporates an optimization process to optimally place phase screens along the optical path before executing the split-step routine. This routine takes in the $C_n^2$ profile over range with the desired intervals between screens. Here, screen intervals may be linear or nonlinear based on user input. Next, TurPy takes in a user-imposed limit for allowable turbulence strength over an interval. If the interval exceeds the allowed Rytov variance, the interval is split until each interval falls within the required limit. If neighboring intervals may sum and remain below the Rytov limit, they are summed. Once roughly optimized, final placement is done through a linear fit to place distances closer to the desired Rytov limit. Finally, each interval is compared against the Fourier optics constraint presented in **Eq. 7**. If the interval remains above this quantity, it is split again until this constraint is achieved. Through this routine, both Fourier optics and phase screen assumptions are maintained for physical accuracy during simulation.

### 2.5 D2NN Training

To train the D2NN, a 0.15 mrad Gaussian beam with a center wavelength of 1064 nm is simulated on a 1024 × 1024 grid with a real-space pixel size of 0.34 mm. During each epoch, the beam propagates through a unique realization of optical turbulence generated by TurPy along a fixed 1500 m Hufnagel-Valley 5/7 path with $\sigma_R^2 = 0.2$. The resultant field is apodized by a 12-inch aperture and passed through a two-mask D2NN, with masks separated by lenses that alternate between real and Fourier domain representations. The lenses are simulated as discrete Fourier transforms following Fourier optics. The processed field is then compared to the free-space result, which is propagated without turbulence or the D2NN, through a mean absolute error loss function, encouraging the D2NN to statistically recover the nominal beam from the turbulence-degraded input and reducing outlier intensity fluctuations.

The D2NN is trained for 20,000 epochs at a learning rate of 0.01 using a one-cycle learning rate schedule to promote convergence. A constant phase initialization is used alongside the high learning rate to promote reproducible exploration of the design space. Training was completed on an Nvidia RTX 4090 GPU in 11 minutes.

## 3. TIME AVERAGED STATISTICS

To verify TurPy, we simulate optical fields and analyze their aggregate behavior against theory. Long-time-averaged statistics are defined by well-established closed-form solutions within atmospheric optics [1], derived from the statistical moments of the mutual coherence function. These moments carry direct physical meaning, such as mean intensity (1st order), long-time broadening (2nd order), and scintillation (4th order). This meaning allows simulated phenomena to be compared against analytical predictions. We analyze the broadening of Gaussian beams and scintillation of plane waves through TurPy following a vertical Hufnagel-Valley 5/7 path (see **Appendix C**) of variable length and ground-level turbulence to verify that induced phase fluctuations follow expected behavior. We first recover the initialized PSD from TurPy to confirm that our routine preserves user-defined turbulence descriptions and determine the number of iterations required to yield statistically stable results in **Sec. 3.1**. We then determine broadening of a Gaussian beam from weak ($\sigma_R^2 \ll 1$) to strong ($\sigma_R^2 \gg 1$) turbulence against known atmospheric models in **Sec. 3.2**, and scintillation of a plane wave across the same turbulence range in **Sec. 3.3**. These results demonstrate TurPy as a high-fidelity optical turbulence surrogate capable of producing statistically consistent and theoretically accurate optical fields for machine learning and synthetic data applications.

## 3.1 Recovery of Power Spectral Density from Phase Screens

The goal of this experiment is to confirm that ensemble-averaging TurPy realizations recovers the user-defined turbulence PSD. Because phase screens are generated through Gaussian-weighted random modulation of the PSD, averaging their Fourier spectra over sufficiently many decorrelated realizations should converge to the input PSD. This experiment additionally determines the averaging threshold required for stable aggregate statistics used in subsequent sections.

TurPy is initialized with a modified Von Karman atmospheric turbulence PSD (**Eq. S17**), and the true PSD is generated from the defined simulation parameters, as shown in **Fig. 4A**. Randomly modulated phase screens are then generated and their instantaneous PSDs averaged, as shown in **Fig. 4C**. The mean square error (MSE) between the true and recovered PSD is monitored over 5000 averaging iterations. As shown in **Fig. 4B**, the MSE decreases rapidly, and the recovered PSD converges to exhibit the correct structure and scaling. The MSE stabilizes at iteration 4000, beyond which individual phase screen contributions are sufficiently averaged to support analysis of aggregate behavior. We adopt 4000 iterations as our averaging threshold for all higher-order statistical experiments that follow.

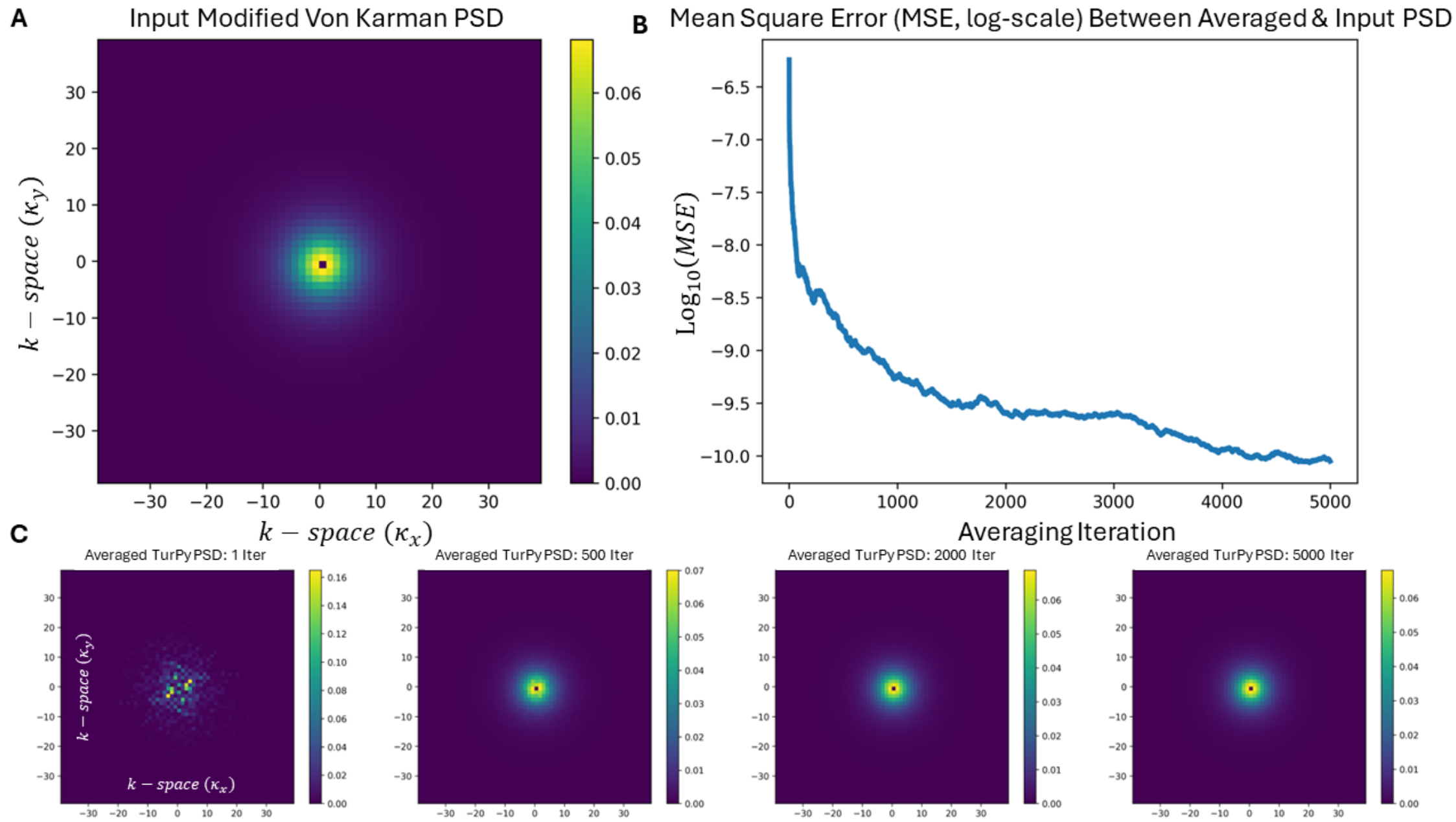

**Fig. 4. Recovery of Turbulence PSD.** (**A**) Input PSD. (**B**) MSE between input and recovery PSD. (**C**) Example recovered PSD after TurPy phase screen averaging.

## 3.2 2nd Order Statistics Validation through Gaussian Beam Broadening

2nd order statistics describe key turbulence-induced beam properties including broadening, wander, angle of arrival, and spatial coherence loss. We analyze Gaussian beam broadening due to its predictable intensity profile recovered after long-time integration. Gaussian beams with a divergence angle of 0.05 mrad and center wavelength of 1064 nm are simulated over paths from 10 m to 5 km, with ground-level turbulence levels (constant $A$ in **Eq. S21**) ranging from $10^{-13}$(low) to $10^{-11}$(very high) m$^{-2/3}$. The long-time broadened radius is estimated from 4000 independent TurPy trials using a 1/e² intensity cutoff.

TurPy closely matches theoretical predictions for the long-time beam waist across all tested conditions. At low to moderate turbulence (**Fig. 5A-B**), the simulated waist agrees with theory to within 99.7%. At high turbulence (**Fig. 5C**), the simulated waist slightly overapproximates the theoretical value, though agreement remains within 2.5% across all conditions. This discrepancy arises because severe wavefront error at long ranges causes the ensemble-averaged profile to converge more slowly to a Gaussian, biasing the 1/e² cutoff upward. Increasing the number of averaging iterations reduces this effect at the cost of simulation time. **Fig. 5D** illustrates this convergence for a moderate turbulence path ($A = 10^{-12}$ m$^{-2/3}$) by comparing the averaged output after 1 and 4000 iterations against the free-space waist (blue circle) and theoretical turbulent waist (red dashed circle). Under-averaged results retain instantaneous wavefront distortions that distort the profile and bias the 1/e² estimate upward.

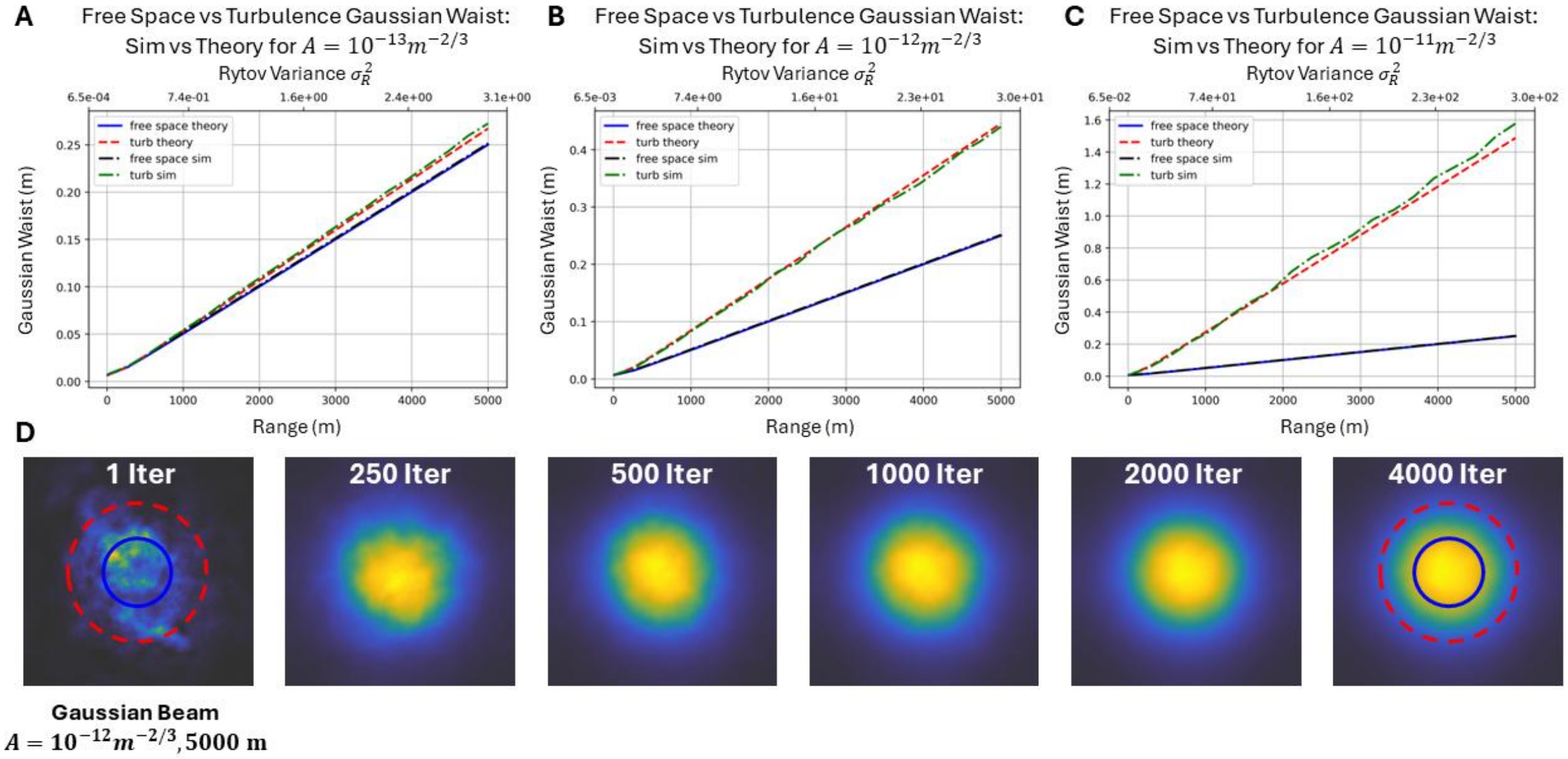

**Fig. 5. 2nd Order Broadening of a Gaussian Beam**. **(A)** TurPy simulated free space and turbulent Gaussian beam size (e.g., waist) versus theory for weak turbulence, **(B)** moderate turbulence, and **(C)** strong turbulence over paths of increasing length. **(D)** Example averaged TurPy outputs to recover an averaged broadened Gaussian for characterization.

### 3.3 4th Order Statistics Validation through Plane Wave Scintillation

4th order statistics describe intensity variance under turbulence and characterize the scintillation of an optical field along a propagation path. To capture scintillation, a plane wave is simulated through TurPy and intensity fluctuations are recorded across 4000 independent trials using Welford's algorithm to compute the scintillation index (SI, **Eq. S25**). The experiment uses a fixed 20 km vertical Hufnagel-Valley 5/7 path with ground-level turbulence ranging from $A = 10^{-17.5}$to $10^{-13.5}$, corresponding to Rytov variances of $\sigma_R^2 = 0.235$to $235.0$. The longer path relative to the broadening experiment allows small phase errors to accumulate sufficient variance for strong turbulence effects, including chaotic speckling, to fully develop. The spatial discretization $\Delta x$is set to the order of the Fried parameter to prevent implicit speckle averaging. Average SI across the output plane is plotted against weak (**Eq. S26**) and strong (**Eq. S27**) scintillation models in **Fig. 6**.

TurPy reproduces the weak-turbulence scintillation scaling up to $\sigma_R^2 \sim 8$, after which the output transitions toward the strong-turbulence saturation model and reaches full saturation by $\sigma_R^2 \sim 100$ (**Fig. 6A**). This result confirms that TurPy captures turbulent behavior across both regimes, including the transition region, supporting physically accurate simulation beyond the limits of either model alone. At $\sigma_R^2 \sim 8$, the scintillation map exhibits high-intensity hotspots driven by underdeveloped caustics rather than homogeneous speckle (**Fig. 6B**). By $\sigma_R^2 = 236$, multiple scattering dominates, and the wavefront is sufficiently randomized to produce well-developed speckle with stable, homogeneous statistics (**Fig. 6C**).

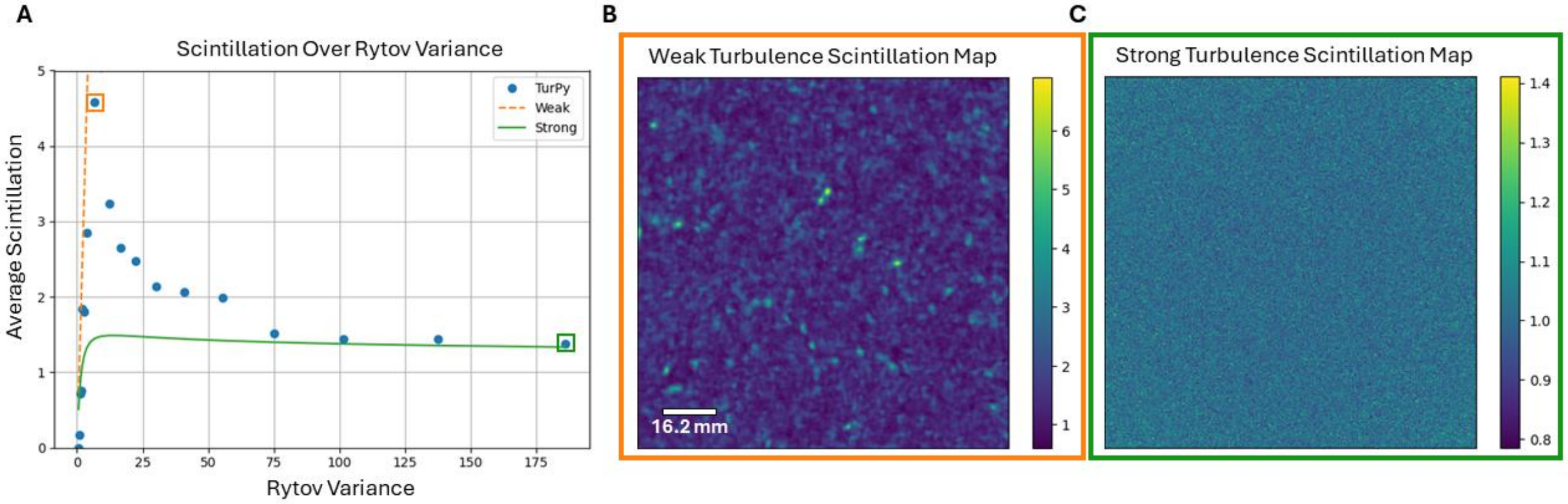

**Fig. 6. 4th Order Scintillation of a Plane Wave. (A)** Average scintillation index from a plane wave passing through a turbulent path with variable strength. **(B)** Example of underdeveloped caustics generated in the weak turbulence regime. **(C)** Example of well-developed speckles generated in the strong turbulence regime.

## 4. TRAINING DIFFRACTIVE DEEP NEURAL NETWORKS FOR ALL-PASSIVE ADAPTIVE OPTICS ALTERNATIVES

To demonstrate TurPy's utility as a machine learning tool, we optimize an all-optical receiver using a trained D2NN to recover optical fields from a weakly turbulent path. This paradigm leverages the statistical inference capacity of D2NNs to passively recover optical signals through turbulence using entirely static trained optics and requiring no moving components or feedback loops. TurPy's differentiable and efficient architecture serves as both the synthetic data generation engine and the gradient calculation tool, enabling end-to-end training of all-optical platforms for applications in turbulence compensation and directed energy.

### 4.1 Training of a D2NN For All-Passive Recovery of Optical Signals Through Turbulence

A 2-mask D2NN is trained to recover a 0.15 mrad Gaussian beam over a 1500 m Hufnagel-Valley 5/7 path with $\sigma_R^2 = 0.22$. Whereas prior work places the D2NN in either the real or Fourier domain, we adopt a dual-domain architecture with one mask in the pupil plane and one in the Fourier plane to jointly learn correction of defocus, modeled through a Fourier-domain transfer function, and turbulence-induced phase error, modeled through real-space pseudo-random modulation. As shown in **Fig. 7A**, the receiver is modified to include the two trained phase masks after the aperture, separated by lenses that induce optical Fourier transforms to realize the dual-domain architecture. The receiver aperture is set to 12 inches but may be scaled as required. The current simulation places the first phase mask at the aperture plane, which may be impractical at full aperture size. Alternatively, a relay system may demagnify the optical field before modulation to alleviate manufacturing constraints.

Once trained, the D2NN exhibits strong phase encoding in the first mask followed by weaker phase encoding in the second, as shown in **Fig. 7B**. The first mask redirects deflected light toward the optical axis by predominantly redirecting off-axis energy to begin reconstructing the nominal beam shape by adopting a near-circular lens-like pattern. The second mask applies lighter phase correction within the Fourier-transformed beam footprint while suppressing artifacts or remaining stray light through its exterior speckled phase structure. To evaluate performance, 4000 independent turbulent paths are generated through TurPy, and the ensemble intensity and scintillation are computed both with and without the D2NN, denoted as naïve and D2NN, respectively.

The D2NN recovers an intensity profile consistent in shape and scale with the free-space Gaussian, confirming that the architecture is performing beam recovery rather than arbitrary redistribution (**Fig. 7C**). Residual speckling artifacts are present in the recovered field, likely due to the constant phase initialization used for reproducibility potentially biasing optimization toward a local minimum. Despite this suboptimal behavior, the D2NN achieves over 58% reduction in scintillation across the beam profile relative to the naïve receiver, confirming that the architecture substantially reduces turbulence-induced intensity fluctuations and stabilizes the beam. Beam scintillation was determined by calculating per-pixel scintillation index and summing scintillation within the 99.9% power cutoff of the average intensity profile to capture scintillation within the bulk of the beam distribution across the turbulent channel. This result demonstrates TurPy's capacity to train novel all-optical architectures for passive turbulence compensation through end-to-end gradient-based optimization.

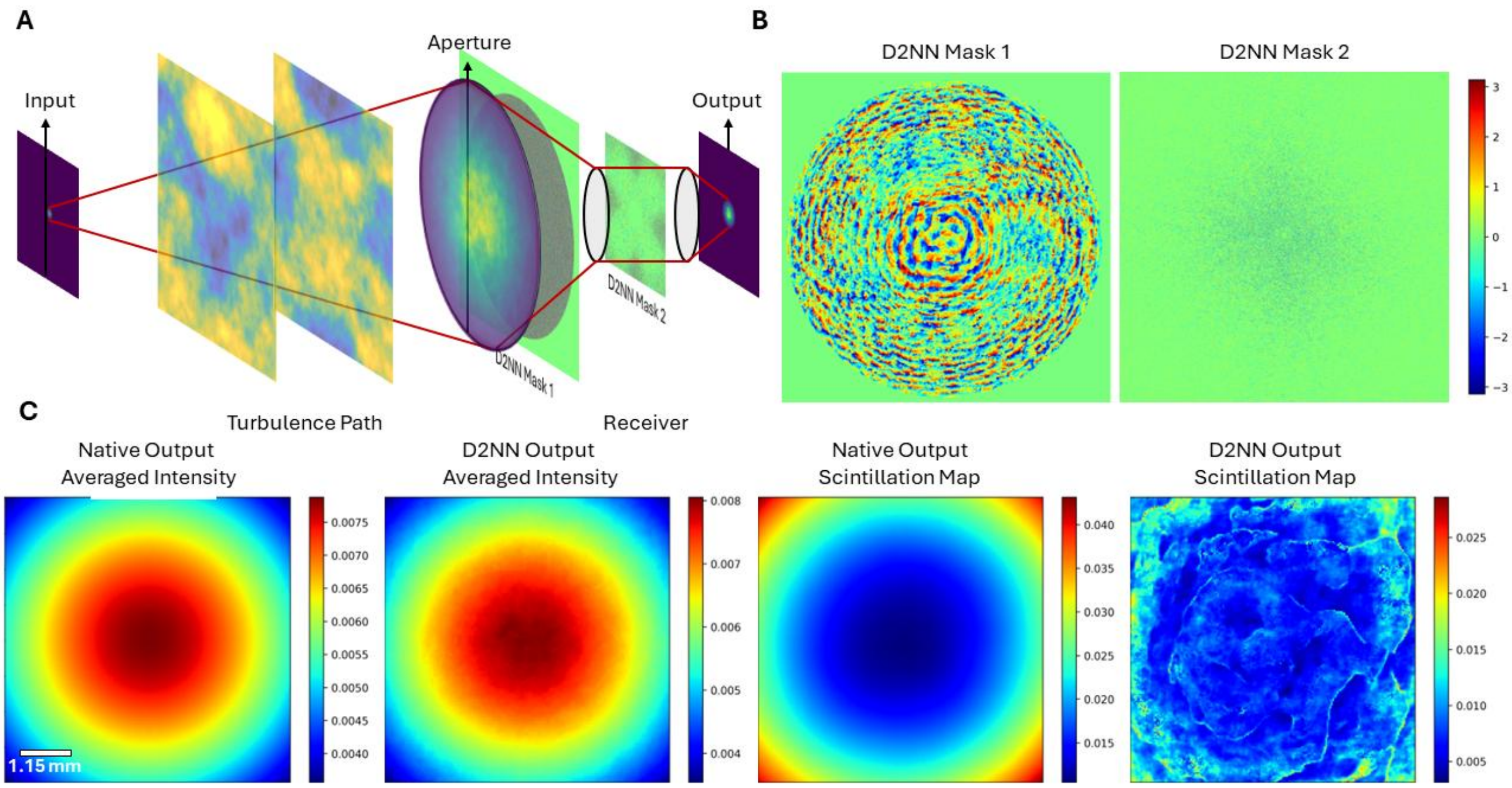

**Fig. 7. Training of a D2NN for Beam Stabilization over Turbulence. (A)** Overview of simulated environment to train D2NN. **(B)** Trained D2NN masks. **(C)** Naïve (e.g., no D2NN) vs D2NN ensemble intensity and scintillation determined over 4000 unique turbulence simulations.

## 5. CONCLUSIONS

This work introduces TurPy, a differentiable, GPU-accelerated synthetic data generation tool for simulating optical fields through turbulence. As a physical simulator, TurPy unifies subharmonic generation and autoregressive temporal evolution to capture outer scale and temporal correlation effects within a single framework. An automated phase screen placement routine enforces both Fourier aliasing constraints and weak-turbulence approximations, ensuring physical validity across user-defined propagation profiles without manual tuning. Validated against atmospheric optics, TurPy reproduces 2nd order Gaussian beam broadening and 4th order plane wave scintillation theory to within 98% accuracy across weak to strong turbulence regimes, confirming its fidelity as a high-quality synthetic testbed. To demonstrate its utility as a gradient-based training platform, TurPy is used to train a dual-domain D2NN for passive recovery of optical fields from a weakly turbulent path. This result achieves a 58% reduction in scintillation compared to an uncompensated receiver to establish TurPy's capacity to optimize all-optical architectures to achieve defined tasks.

As a machine learning tool, TurPy centralizes differentiable wave optics, GPU acceleration, and configurable turbulence descriptions into a unified package, lowering the barrier for end-to-end optical system design across research communities. While validated here against atmospheric turbulence, oceanic and biological propagation environments exhibit well-characterized PSDs (see **Appendix B**) and are supported by TurPy's PSD-parameterized architecture with minimal modification. Domain-specific path-integrated quantities, such as the Rytov variance, may adapt to these settings, requiring only updates to the governing PSD and scaling constants [21].

TurPy's multi-screen architecture is inherently anisoplanatic, meaning that field decorrelation across the propagation path is captured without additional modification. This property positions TurPy for imaging applications, where the incoherent point spread function may be simulated and applied patch-wise through local convolution to generate spatially decorrelated image degradation consistent with real turbulent conditions [22]. Extending support for polychromatic and partially coherent illumination [12] would further push TurPy's capabilities to passive imaging systems and scenarios where source coherence cannot be assumed, such as remote sensing applications.

These results position TurPy as a general-purpose synthetic data generation and end-to-end optimization platform spanning atmospheric, oceanic, and biological domains. As differentiable optical design continues to mature, TurPy will bridge physics-accurate simulation with gradient-based learning to push robust, deployable optical systems in turbulence applications.

## ACKNOWLEDGEMENTS

We would like to thank the Georgia Tech Research Institute Independent Research and Development (IRAD) for providing support for this project. We would like to thank Nathan Meraz for his insightful comments and conversation throughout developing TurPy. Thank you to Chico Andre Olaguer at De La Salle University – Manilla for aiding in reviewing the D2NN training loop.

## CODE AVAILABILITY

The code used to generate the results is available in a Github repository: https://github.com/joeg18/TurPy/

**Appendix A: Comparison of TurPy to Other Simulation Engines**

Here, TurPy is compared to other python-based open-source wave optics simulation engines adaptable to turbulence to ground its functionality and relative maturity to motivate its utility as an end-to-end simulation tool. This analysis highlights Python capabilities as a leading language for data-driven development due to its access to comprehensive tensor-based libraries such as TensorFlow and PyTorch. This exercise is used to highlight the strengths of TurPy but also emphasize areas for further development as it is matured towards a standard tool for end-to-end applications. In this table (see **Table 1**), ✓ indicates full explicit support, ~ indicates partial or indirect support, and ✗ indicates limited to no support. Partially or indirect support categories are discussed after **Table** 1. This conversation is scoped to other available open-source tools relevant to simulating turbulence and, while potentially not exhaustive, is designed to capture major advancements to define the general trends and emerging functionality in the field. This comparison will include: AOTools [23], HCIPy [24], WavePy [25], WavePy 2 [26], POPPY [27], Diffractio [28], Torch Optics [29], and PADO [30]. The first six are direct turbulence engines with a focusing on atmospheric statistics while the last two are wave optics engines with a focus on differentiability. While it is recognized that some of the properties compared below can be enabled through external Python libraries, such as Jax for autodifferentiation or Reikna for GPU-accleration, this comparison focuses on out-of-the-box capabilities achievable by a researcher with minimal modification.

| | **AOTools** | **HCIPy** | **WavePy** | **WavePy 2** | **POPPY** | **Diffractio** | **Torch Optics** | **PADO** | **TurPy** |
|---|---|---|---|---|---|---|---|---|---|
| User-defined turbulence engine | ✓ | ✓ | ✓ | ✓ | ~ | ✓ | ✗ | ✗ | ✓ |
| Subharmonics | ✓ | ✓ | ✓ | ✓ | ✗ | ✗ | ✗ | ✗ | ✓ |
| Temporal Correlation | ~ | ~ | ~ | ~ | ✗ | ✗ | ✗ | ✗ | ✓ |
| Strong Turbulence Support | ✗ | ✗ | ✗ | ✗ | ✗ | ✗ | ✗ | ✗ | ~ |
| Anisoplanatic propagation | ✓ | ✓ | ✓ | ✓ | ✗ | ✓ | ✗ | ✗ | ✓ |
| Automated Phase Screen Placement | ✗ | ✗ | ~ | ~ | ✗ | ✗ | ✗ | ✗ | ✓ |
| Multi-Domain PSD Support | ✗ | ✗ | ~ | ~ | ✗ | ✗ | ✗ | ✗ | ~ |
| GPU-Accelerated | ✓ | ✓ | ✗ | ✗ | ✓ | ✓ | ✓ | ✓ | ✓ |
| Differentiable / End-to-End Support | ✗ | ✗ | ✗ | ✗ | ✗ | ✗ | ✓ | ✓ | ✓ |
| Partially Coherent / Polychromatic | ✗ | ~ | ✗ | ✗ | ~ | ✓ | ✓ | ✗ | ✗ |

**Table 1. Comparison of open-source python wave optics simulators extendible to turbulence.**

For partial functionality, ~ indicates:

- User-defined turbulence engine – generation of a single-phase screen at the pupil of the simulated optical platform and cannot capture multi-phase errors along a path (POPPY).
- Temporal Correlation – Translational only temporal correlation used by generating extended phase screens and cropping operators (AOTools, HCIPy, WavePy, WavePy 2).

- Strong Turbulence – Demonstrates indirectly control strong turbulence transition through direct subdivision of a turbulent path with weak intervals, however does not include explicit strong models, such as extended Huygen-Fresnel (TurPy).
- Automated Phase Screen Placement – Optimizes split-step interval based on Fourier constraints without constraining phase error.
- Multi-Domain PSD Support – Enables non-Kolmogorov atmosphere phase screens with control over power law scaling, but does not allow explicit definition of PSDs from other domains.
- Partially Coherent / Polychromatic – Allow for multi-wavelength simulations with rescaled phase screens versus full broadband calculation.

**Appendix B: Derivation of Phase Screen Scaling Laws**

To generate a phase screen, we will utilize a common approach of deriving a power spectral density (PSD) function to describe the spatial correlation of turbulence across simulated scales, then randomly weight the path-integrated phase factor with a unit Gaussian random variable to yield a representative instance [1].

To simplify the scope of simulating optical turbulence, we begin by making a number of standard assumptions in our model: 1.) we model the effect of optical turbulence on a coherent field. 2.) Optical turbulence may be decomposed into zero-mean index fluctuations surrounding a mean index. 3.) Optical turbulence statistics are statistically homogenous (e.g., stationary) and rotationally invariant (e.g., isotropic). 4.) The temporal evolution of optical turbulence is slow compared to the speed of light (e.g., frozen flow) [1]. In other words, we are simulating complex monochromatic optical fields through turbulence, under which we assume the index variations are centered and are locally consistent, large, and slowly changing when compared to the wavelength and oscillations of light.

We first begin by defining our index of refraction by:

$$n(r) = n_o + n'(r) \tag{S1}$$

where $n(r)$ is the spatially varying refractive index in spherical coordinates, $n_o$ is the mean index and $n'(r)$ is the zero-mean, statistically homogeneous random field in 3D space.

Next, we describe the index of refraction structure function $D_n(\rho)$, which describes the statistical variation, or “roughness”, of the index as a function of separation between two points separated by $\rho$, given by:

$$D_n(\rho) = \langle [n'(r+\rho) - n'(r)]^2 \rangle \tag{S2}$$

where $D_n(\rho)$ is assume to be radially and translationally invariant due to the isotropic and stationary assumptions of turbulence. In practice, $D_n(\rho)$ is closely related to the index of refraction structure constant, $C_n^2 = D_n(\rho)/\rho^{2/3}$, which described the strength of turbulence over a spatial interval and will be used later on as a general description of turbulence.

Empirically, the refractive index structure function is observed to comprise of an inertial range, where the turbulent energy cascades from large scales of motion to small scales in accordance with a power law, and non-inertial region wherein turbulent energy dissipates [31]. As such, models define an inner scale ($l_o$) and outer scale ($L_0$) of turbulence to bound the inertial region with a general structure of:

$$D_n(\rho) \propto \rho^{\beta} \tag{S3}$$

Where $\beta$ is a media dependent power law factor.

We may modify the definition of the structure function in **Eq. S3** to describe it in terms of a spatial index correlation function, $B_n(\rho)$ by expanding:

$$D_n(\rho) = \langle n'(r+\rho)^2 + n'(r)^2 - 2n'(r+\rho)n'(r) \rangle = 2[B_n(0) - B_n(\rho)] \tag{S4}$$

which relates to a PSD through the Wierner-Khinchin theorem:

$$B_n(\rho) = \int \Phi_n(\kappa) e^{i\kappa\cdot\rho} \frac{d^3\kappa}{(2\pi)^3} \tag{S5}$$

where $\Phi_n(\kappa)$ is the PSD and $\kappa$ is the radial wavevector coordinate. By plugging **Eq. S4** into **Eq. S5** yields:

$$D_n(\rho) = 2\int [1 - e^{ik\cdot\rho}]\Phi_n(\kappa) \frac{d^3\kappa}{(2\pi)^3} \tag{S6}$$

Next, we reformulate **Eq. S6** in spherical coordinates (e.g., $d\kappa^3 = \kappa^2 sin\theta d\phi d\kappa d\theta d\phi$) and impose that $k \cdot \rho = k\rho cos\theta$. Plugging in yields:

$$D_n(\rho) = \frac{2}{(2\pi)^3}\int_0^\infty \Phi_n(\kappa) \times \left(4\pi - \int_0^{2\pi} d\phi \int_0^{\phi} (e^{ik\rho cos\theta}) sin\theta \, d\theta\right)\kappa^2 d\kappa \tag{S7}$$

which can be further reduced by solving the integral over solid angle to yield , $\int_0^{2\pi} d\phi \int_0^{\phi} (e^{ik\rho cos\theta}) sin\theta \, d\theta = 4\pi \frac{\sin(\kappa\rho)}{\kappa\rho}$. Using a similar assumption to **Eq. S3** to enforce that $\Phi_n(\kappa) \propto \kappa^{-\alpha}$ within the inertial regime by the reciprocal nature of the Fourier transform and collecting all constants into an arbitrary value, $c$, we may yield:

$$D_n(\rho) \sim c\int_0^\infty \kappa^{2-\alpha}\left[1 - \frac{\sin(\kappa\rho)}{\kappa\rho}\right] d\kappa \tag{S8}$$

which is a modified Hankel Transform. To solve, we perform a u-substitution of $u = \kappa \cdot \rho$ to yield:

$$D_n(\rho) = c\rho^{3-\alpha}\int_0^\infty u^{2-\alpha}\left[1 - \frac{\sin(u)}{u}\right] du \tag{S9}$$

where the remaining integrand integrates to a constant under light assumptions [32]. As such, we may approximate the 3D PSD up to a constant error by:

$$D_n(\rho) \propto c\rho^{3-\alpha} \tag{S10}$$

By comparing the power laws between **Eq. S3** and **Eq. S10**, we can determine the corresponding power law factor in real and k-space by:

$$\alpha = \beta + 3 \tag{S11}$$

As such, if we understand the scales of turbulence in one domain, we may accurately describe its scaling in the reciprocal domain. We present a number of complimentary $\alpha$ and $\beta$ factors in **Table S1**.

| **Media** | $\boldsymbol{\beta}$ | $\boldsymbol{\alpha}$ |
|---|---|---|
| Atmosphere (Komolgorov) [6] | 2/3 | 11/3 |
| Atmosphere (General) [33] | 0-1 | 3-4 |
| Oceanic [2] | 2/3 – 2 | 11/3 – 15/3 |
| Biological Tissue [3] | 1.28 – 1.41 | 4.28 – 4.41 |

**Table S1**: Power law scaling of characterized media

Now that we characterized the general power law scaling of our media, we may write our PSD in the general form:

$$\Phi_n(\kappa) \propto \kappa^{-\alpha} = \kappa^{-(\beta+3)} \tag{S12}$$

Once we define the PSD, we may project the path-dependent index fluctuations into a 2D representative phase screen through:

$$\varphi_n(r_\perp;\Delta z) = k\int_0^{\Delta z} n'(r_\perp, z)dz \quad \text{S13}$$

where $r_\perp$ is the transpose component of $r$, z is the axial direction, $k$ is the wavenumber and $n'(r_\perp, z)$ is the random index fluctuations in cylindrical coordinates. As our index fluctuations are isotropic, we may assume that this cylindrical reparameterization follows the same power law as $n'(r)$.

Using a similar strategy as **Eq. S5**, we may relate our described phase screen to a spatial phase correlation and relate it to the path integrated 3D refractive index PSD, yielding:

$$\Phi_\varphi(\kappa_\perp;\Delta z) = k^2\int_0^{\Delta z} \Phi_{\mathrm{n}}(\kappa_\perp, z)dz \quad \text{S14}$$

Where $\kappa_\perp$ is the transverse wavevector, $\Phi_\varphi$ is the 2D PSD of our phase screen and $\Phi_{\mathrm{n}}$ is our 3D refractive index PSD.

By similar logic to **Eq. 5**, $\Phi_{\mathrm{n}}(\kappa_\perp, z)$ may be described as a power law quantity to re-express **Eq. 7** as:

$$\Phi_\varphi(\kappa_\perp; z) \propto k^2\Delta z\kappa_\perp^{-\alpha} \quad \text{S15}$$

This formulation yields a simple yet powerful approach to roughly characterize a phase screen for application across domains, as TurPy only requires information on the wavelength and power law to estimate the phase screen PSD for a given application. Additionally, when turbulence remains uncorrelated over the length-scale of our step size (e.g., the Markov approximation), each phase screen contributes independently to the total phase variance along the path to support physical accuracy.

Numerically, we may adapt **Eq. S15** to yield pseudo-random phase screens by using the path-integrated phase PSD as a standard deviation of a complex Gaussian random variable. For completeness, we reintroduce our cartesian coordinates from the beginning to recognize that we will simulate our projected phase screen on 2D synthetic grids on which we will implicitly enforce the radial symmetry from above. As a result, we yield:

$$\varphi_n(x, y;\Delta z) = \mathfrak{I}^{-1}\{\sqrt{\Phi_\varphi(\kappa_x, \kappa_y;\Delta z)\Delta\kappa_x\Delta\kappa_y} \times \mathrm{N_c}(0,1; N_x, N_y)\} \quad \text{S16}$$

Where x, y are now simulated grids of size $N_x, N_y$, $\kappa_x, \kappa_y$ is our simulated $\kappa -$ space with discretization $\Delta\kappa_x, \Delta\kappa_y$, the term $\Delta\kappa_x\Delta\kappa_y$ preserves power in accordance with Parseval's theorem, and $\mathrm{N_c}(0,1;\ N_x, N_y)$ is a standard zero-mean unit-variance complex Gaussian random process of the size of our simulated grid. After randomly weighting the PSD, a phase screen in real space may be realized by taking the inverse DFT. Once generated, the phase screens are ready for use in split-step simulations through **Eq. 1**. As with many phase screen simulators, we remove the mean from the resultant phase screen to remove piston in the final result [6].

We may adapt this formula to different media through characterized power law behavior, as presented in **Table S1**. In addition, we express the phase PSD precisely for atmospheric turbulence in **Appendix B**, as its behavior is well understood and will be used to validate TurPy within this work.

**Appendix C: Phase PSD & Statistical Quantities for Atmospheric Turbulence**

We validate TurPy against atmospheric turbulence as its well-characterize and rigorous models serve as a motivating target for TurPy to replicate. To define atmospheric turbulence, we model the path-integrated phase fluctuations through the Modified Von Karman phase PSD [1] both within and outside the inner and outer scale in accordance with:

$$\Phi_{\varphi}(\kappa_{\perp};\Delta z) = 0.023 r_o^{-5/3} \frac{e^{-\frac{\kappa_{\perp}^2}{\kappa_m^2}}}{(\kappa_{\perp}^2+\kappa_0^2)^{11/6}} \tag{S17}$$

Where $\kappa_m = \frac{2\pi}{l_0}$, is the wavevector associated with the inner scale of turbulence, $\kappa_0 = \frac{2\pi}{L_0}$ is the wavevector associated with the outer scale of turbulence and $r_0$ is the Fried Parameter, which measures the spatial coherence length of an optical field passing through turbulence as has the form:

$$\mathrm{r_o} = \left[0.423k^2 \int_0^{\Delta z} C_n^2(z)\right]^{-3/5} \tag{S18}$$

Where $C_n^2(z) = D_n(\rho)/\rho^{2/3}$ is the refractive index structure constant and describes the strength of optical turbulence irrespective of scale (within the inertial subrange). Additionally, the Fried Parameter holds a close physical relationship to the Rytov variance, $\sigma_R^2$, or a path integrated quantity of turbulence strength over an integral of the form:

$$\sigma_{\mathrm{R}}^2 = 3.27k^{7/6}\Delta z^{5/6} \int_0^{\Delta z} C_n^2(z)\left(1-\frac{z}{\Delta z}\right)^{5/3} dz \tag{S19}$$

In general, a $\sigma_R^2 < 1$ describes a weakly turbulence path and $\sigma_R^2 \gg 1$ describes a strong turbulence path and traditionally the phase screen split-step formulation remains valid for intervals with $\sigma_R^2 < 1$.

When $\sigma_R^2$ is set to 1, we may plug **Eq. S19** into **Eq. S18** to calculate the maximum allowable Fried Parameter under weak approximations for a single-phase screen as:

$$\mathrm{r_o} = \left[\frac{0.129k^{\frac{5}{6}}}{\Delta z^{5/6}} \frac{1}{\frac{\int_0^{\Delta z} C_n^2(z)\left(1-\frac{z}{\Delta z}\right)^{\frac{5}{3}} dz}{\int_0^{\Delta z} C_n^2(z)}}\right]^{-3/5} \tag{S20}$$

We use this quantity in **Sec. 2.4** to calculate our optimally placed phase screens.

In general, the refractive index structure constant decreases with height, leading to a Fried Parameters (and therefore total amount of phase accumulation), which depends on the orientation of light propagation through the atmosphere. In this work, we assume a vertical light-path with a Hufnagel-Valley 5/7 profile, which describes a height dependent $C_n^2$ in a deterministic equation tunable through a single ground level constant, $A$, and takes the form:

$$C_n^2(z) = 0.00594\left(\frac{V}{27}\right)^2 (10^{-5}\times z)^{10} e^{\frac{-z}{1000}} + (2.7\times 10^{-16}) e^{\frac{-z}{1500}} + A e^{\frac{-z}{100}} \tag{S21}$$

Where $V$ describes an assumed wind speed in m/s$^2$.

To confirm TurPy as a high-fidelity synthetic tool, we seek to match simulated characteristics of optical fields through TurPy's split step model to validated theory. We specifically look to replicate 2$^{nd}$ order (e.g., broadening) behavior of Gaussian beams and 4$^{th}$ order (e.g., scintillation) of plane waves using TurPy when compared to established theory [1].

A free space (e.g., no turbulence) Gaussian beam is parameterized by its divergence angle, $\theta_0$, and fundamental waist, $w_0$, and exhibits an 2D intensity distribution of the form:

$$I(x,y) = I_o e^{-2\left(\frac{x^2+y^2}{w_o^2}\right)} \quad \text{S22}$$

Where $I_o$ is the intensity scaling factor and the waist and divergence angle are related through $w_0 = \frac{\theta_0}{\pi\lambda}$ where $\lambda$ is the beam center wavelength.

After propagating by a distance, $z$, the intensity distribution becomes:

$$I(x,y;z) = \frac{I_o w_0^2}{W(z)^2} e^{-2\left(\frac{x^2+y^2}{W(z)}\right)} \quad \text{S23}$$

Where $W(z)$ is the free space radius, which takes the form $W(z) = \sqrt{w_o^2 + (\theta_0 z)^2}$.

Under turbulence, it may be shown [1] that the broadened field retains a Gaussian shape with an additional turbulence broadening factor, $\xi = 1.33(\sigma_R^2)^{\frac{3}{5}}\Lambda$, where $\sigma_R^2$ is a path integrated scalar metric of turbulence known as the Rytov variance, and $\Lambda = \frac{2z}{\frac{2\pi}{\lambda}W(z)^2}$ is the Fresnel parameter of a Gaussian beam. Together, the long-time turbulence-broadened radius, $W_{LT}(z)$ takes the form:

$$W_{LT}(z) = W(z)\sqrt{(1+\xi)} \quad \text{S24}$$

And may be directly substitute $W(z)$ for $W_{LT}(z)$ in **Eq. S23**.

Next, we explore the scintillation of a plane wave. Scintillation index (SI) is defined as rapid fluctuations in intensity caused by an optical field passing through turbulence. In general, scintillation is considered a 4th order effect, and is related to the standard deviation of the intensity, $\sigma_I^2$ and average intensity, $I_{av}$ by:

$$SI = \frac{\sigma_I^2}{I_{av}^2} \quad \text{S25}$$

For plane waves, scintillation follows differing descriptions for weak (e.g., $\sigma_R^2 < 1$ and strong turbulence models (e.g. $\sigma_R^2 \gg 1$). For weak turbulence, plane wave SI scales linearly with the Rytov variance:

$$SI_{weak} = 1.23\sigma_R^2 \quad \text{S26}$$

However, at strong turbulence, phase coherence strongly degrades and chaotic scattering dominate leading to a plateauing of the SI in accordance with:

$$SI_{weak} = \exp\left(\frac{0.49\sigma_R^2}{\left(1+1.11\sigma_R^{2\frac{6}{5}}\right)} + \frac{0.51\sigma_R^2}{\left(1+0.69\sigma_R^{2\frac{6}{5}}\right)^{\frac{5}{6}}}\right) - 1 \quad \text{S27}$$